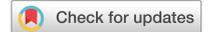

# OPEN    Effects of multimodal explanations for autonomous driving on driving performance, cognitive load, expertise, confidence, and trust

Robert Kaufman[1✉], Jean Costa[2] & Everlyne Kimani[2]

Advances in autonomous driving provide an opportunity for AI-assisted driving instruction that directly addresses the critical need for human driving improvement. How should an AI instructor convey information to promote learning? In a pre-post experiment (n = 41), we tested the impact of an AI Coach's explanatory communications modeled after performance driving expert instructions. Participants were divided into four (4) groups to assess two (2) dimensions of the AI coach's explanations: information type ('what' and 'why'-type explanations) and presentation modality (auditory and visual). We compare how different explanatory techniques impact driving performance, cognitive load, confidence, expertise, and trust via observational learning. Through interview, we delineate participant learning processes. Results show AI coaching can effectively teach performance driving skills to novices. We find the type and modality of information influences performance outcomes. Differences in how successfully participants learned are attributed to how information directs attention, mitigates uncertainty, and influences overload experienced by participants. Results suggest efficient, modality-appropriate explanations should be opted for when designing effective HMI communications that can instruct without overwhelming. Further, results support the need to align communications with human learning and cognitive processes. We provide eight design implications for future autonomous vehicle HMI and AI coach design.

Recent years have seen vast improvements in autonomous driving technology, with on-road vehicle testing currently being conducted in major cities around the world. The proposed benefits of autonomous vehicles (AVs) include increases in driving safety and efficiency while reducing driving infractions, traffic, and passenger stress[1]. Though AVs may have a bright future, real-world deployment is hindered by a number of technological, infrastructural, and human-interaction roadblocks that are likely to take decades to solve. Meanwhile, the National Highway Traffic Safety Administration (NHTSA) estimates that in the United States alone, there are over 6 million car crashes each year, resulting in over 2.5 million injuries, 40,000 deaths, and $340 billion in damages[2,3]. Of these, it is estimated that 94% of vehicle crashes are due to human error[4]. Therefore, there is a large and pressing demand for technologies that may improve human driving ability. In this study, we seek to test a novel use of autonomous vehicle technology: can AVs teach humans to be better drivers?

We propose that augmenting driver learning by observing an AI driving coach may help address the glaring need for human driving improvement. To test this concept, we conducted a mixed-methods driving simulator experiment in which participants observed one of four AI coaches providing different types of explanatory instructions. We evaluate how the AI coaching sessions impact learning outcomes such as participant-driving performance, cognitive load, confidence, expertise, and trust.

We leverage the domain of performance or "race track" driving to test study objectives. Performance driving is more challenging but very related to everyday driving, and it allows driving skills and knowledge to be built and tested more objectively. The goal of performance driving is to drive as fast as possible on a given track, maximizing the vehicle's limits while minimizing the distance needed to travel[5]. Many performance driving skills directly translate to real-world driving contexts—such as improvements to vehicle handling and situational awareness[6,7]—and thus, it is an appropriate proxy to study driving skill learning. Testing the potential of an AI driving coach in the context of performance driving has several major benefits over everyday driving. First, performance driving has specifically defined parameters for success so we can objectively measure the effectiveness of our AI coach in a controlled environment. Next, it is a driving task many people are unfamiliar with, and thus we are able to test the potential of an AI driving coach on true novices. This helps maintain a consistent knowledge and skill baseline for our study sample, providing further consistency. Lastly, by testing our AI coach on learning an extreme and challenging driving task, we hope to derive insights that can be generalized to even the most intense real-world driving situations.

Though novel in the realm of autonomous driving, learning from AI, particularly by means of AI explanations, is not a new concept[8]. Learning from AI is a rapidly expanding area of interest, particularly given the proliferation of AI-based large language model tools such as ChatGPT[9,10]. In domains such as medicine, explainable AI-based systems—for example, image classification systems for radiology—have been shown to help physicians learn new techniques to identify pathologies[11–13]. These systems present justifications in the form of 'explanations' which provide a rationale for a system's decisions. Specific cases where human-interpretable AI explanations of decisions are produced have been singled out as especially helpful for learning and improvement[14–16], particularly in cases where AI explanations are based on the explanations of human experts[17,18]. AI has the ability to advance educational techniques in many other domains, from second language learning[19] to programming[20]. Measuring the impact of interacting with AI systems on learning outcomes can be a challenge, and we can differentiate between explicitly testing learning via knowledge tests and operationalizing learning more functionally via outcomes like task performance[21]. For the present study, we measure learning in both ways. The main learning outcomes assess changes in driving performance before and after exposure to the AI coach; secondarily, we assess knowledge via quiz. In this way, we are able to measure the learning impact of the AI Coach directly from multiple perspectives.

In the context of autonomous driving, we propose that observing an AI driving coach may be an effective way to transfer knowledge and teach critical driving skills. There have been a number of studies on human-machine interfaces (HMIs) for AVs, which often take the form of Heads-Up-Displays (HUDs) aimed at building transparency, accountability, and trust[22,23]. Though highly relevant, these studies have majorly focused on non-learning areas of the driver experience, and none to our knowledge have tested the impact of HMI elements on driver learning specifically. For example, in the large corpus of work on human-AV trust formation, Morra et al. found that showing more informative displays increased participant willingness to test a real AV[24], Ruijten et al found that displays mimicking human explanations increases trust[25], and Koo et al. found that including why-type information in display improves trust[26]. Other HMI studies have emphasize the complexity of trust formation, proposing holistic and multi-factor approaches to designing trustworthy interfaces[27–29]. Beyond trust, Schartmüller et al. assessed display modality differences in driver comprehension and multi-tasking, finding that certain interfaces could reduce workload better than others[30]. HMI experiments aimed at building situational awareness with AVs have shown improved awareness with display elements that allow a person to comprehend of why or why-not (contrastive) a vehicle is taking an action[31–33]. Though these prior studies support vehicle HMIs as a promising avenue for coaching applications, how these may apply in instructional applications is currently unexplored. The work presented here seeks to contribute to the knowledge gap that exists between HMI design research and learning communities focused on optimal coaching techniques.

We focus on the role an AI Coach can play in improving a person's driving ability—using explanatory communications modeled after the instructions of human driving experts. As the most common in-car method for introductory performance driving instruction is to begin with observational learning (i.e., a passenger observing an expert driver), our explanatory communications are framed within this observational learning context. Using a state-of-the-art, full-motion driving simulator, we explore both if an AI coach's instructions can improve a novice's driving performance more than observation alone and, if so, what explanatory methods may be best.

Inspired by explanations given by expert driving instructors, we explore the role that HMIs may play in performance driving instruction. To design an effective HMI, it is important to determine both the type of *informational content* that should be conveyed as well as the *presentation modality* of the information. This is especially important in the context of HMIs for safety-critical and cognitively demanding tasks like driving[34]. Studies have shown mixed results on the impact of modality of information presentation. Some studies claim visual techniques should be preferred[30,35], while others suggest auditory feedback as a better strategy[36,37]. Impact of the type and content of information presented also shows mixed results in terms of driver performance and preference, including differences between 'what is happening' or 'why it is happening' information[26,32]. These explanation attributes have not been explored in the domain of AI coaching for driving instruction.

To address these gaps of knowledge, this study uses a mixed-methods approach to assess study outcomes. Participants were randomized into one of four AI coaching groups differing by the type and presentation modality of the information they presented. Before and after observation, measures were taken to compare changes in participant driving performance, cognitive load, confidence, expertise, and trust. Each of these factors have been highlighted as important for the development and adoption of HMIs and explainable AI systems more generally. Interviews were conducted to contextualize findings within the larger learning context, including building a more general understanding on concerns with AVs and how these may be mitigated. This broader view of AVs can help illuminate additional roadblocks that must be addressed for successful future adoption of AI driving coaches.

**Our research questions are as follows:**

1. What are the pre-post impacts of observing an AI performance driving coach compared to a pure observation (no explanation) control, including the impacts of explanation information type and presentation modality?
2. What is the process of a novice learning performance driving, and how can an AI Driving Coach facilitate this learning?

3. What concerns do novices have in general about the deployment of AV technology, and how can these further inform future AV HMI design?

Results from this study support the premise of AI coaching as a promising method for driving instruction and lead to important considerations for future HMI design.

**Specifically, we contribute:**

1. A novel assessment of the impact of observing an AI Coach on performance-driving ability, cognitive load, confidence, expertise, and trust.
2. Insight into the impact of information type ('what' and 'what + why') and information presentation modality (visual and auditory) on the process of learning performance driving.
3. Eight design insights to inform the creation of future human-centered HMIs for driving and AI driving coaches.

## Method

To address our research questions, participants were divided into one of four experimental conditions differing in the *information type* and information *presentation modality* given by the AI coach. For information type, we test two layers of information: (1) 'What' information provides descriptors of where the vehicle should drive; (2) 'Why' information explains the rationale behind why that position and movement is optimal, and is meant to build conceptual understanding. Some participants received just 'what' information, while others received both 'what' and 'why' information in combination. With this manipulation, we seek to build upon prior work on the impact of information level on performance, preference, and trust[26]. For information modality, we manipulated whether 'what' explanations are presented auditorily or in the form of a visual racing line projected on the track; These two additional conditions aim to provide clarification on the efficacy of visual and auditory information at conveying information about a vehicle's behavior.

We evaluate changes in driving performance from before observing the AI coaching condition to after observation. One crucial concept of learning performance driving focuses on vehicle positioning, which is a fundamental skill for novices. In performance driving, the optimal position follows a path on the track called the 'racing line', which minimizes time cost around corners and allows the driver to move as quickly as possible[38,39]. As the AI coach's 'what' instructions primarily focused on the racing line, the main outcome of our study is how well participants positioned themselves while driving. Other performance measures include lap time, speed, and acceleration; these were addressed only in explanations that included a 'why' component.

Additional secondary outcomes were addressed via interview and questionnaire. These include impacts on trust, knowledge and expertise, driving confidence, and feedback related to how helpful and effective the AI coach was at facilitating the participants' learning process.

A large corpus of prior work suggests that a major drawback of mid-drive communication is the potential for information overload and high cognitive demand[34]. We hypothesized that observational learning would provide the opportunity to transfer knowledge to a novice while avoiding cognitive overload, as a participant can pay attention to explanations without task switching between learning and driving decision-making. To further explore this phenomenon, we measure cognitive load for each of our study conditions.

### Participants

A total of 41 participants were recruited for the study and completed all study procedures. Participants were all novices in the performance driving domain but were otherwise licensed to drive with at least 5 years of everyday driving experience. This ensured that participants had the baseline motor skills necessary to begin performance driving instruction.

### Performance driving simulation

All driving-related tasks took place in a state-of-the-art, highly immersive, full-motion driving simulator to ensure realism. This included a 270-degree wrap-around screen, a full vehicle cabin with a closely calibrated steering wheel and pedals, sound effects, wind effects, and cabin movement mimicking the feeling of a vehicle moving on the track (Fig. 1). The simulator has a mounted tablet so that participants can answer survey questions between drives without leaving the simulation, ensuring real-time responses and preserving immersiveness. The specific racing context chosen was a highly accurate simulation of the professional driving course Thunderhill Raceway in Willows, California, USA.

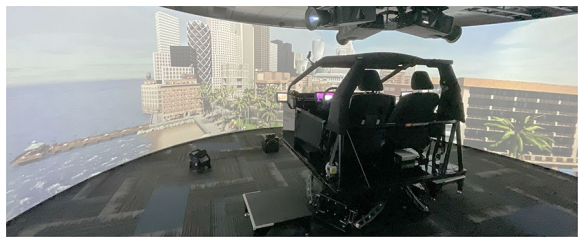

**Figure 1.** Full-motion driving simulator.

During the AI coaching observation sessions, the vehicle drove autonomously with no user input. The specific model used to control the vehicle was a reinforcement learning agent trained using the DSAC algorithm[40], with observations and rewards tuned for Thunderhill. The agent's policy was optimized for the physical features and limitations of the vehicle and the geometry of the track[41]. This model was also used to compute the "ideal" racing line for calculating performance.

### AI coach explanations

During the AI coach observation sessions, explanatory instructions were provided to all participants except the control group. Auditory explanations were presented via an in-cabin speaker. These were produced using Amazon Polly Text-to-Speech[42]. Visual explanations were projected directly onto the track in the simulated environment.

We modeled the AI coach's explanations off instructions given by four real performance driving experts and instructors who all had real-world experience driving and instructing on the real-world Thunderhill Raceway. First, two expert performance drivers were ethnographically observed giving real-world driving instruction at Thunderhill Raceway via several hours of video. Next, this was paired with think-aloud [cite] and interview sessions with two additional expert performance drivers as they drove the same simulated Thunderhill track used in the experiment. This qualitative assessment was used to ground and justify the specific explanations given by the AI coach to ensure accuracy. Descriptions given by the experts also helped scaffold the learning procedure, which we have described in detail later in this paper.

We found that the most common in-car method for introductory performance driving lessons began with observational learning, specifically when an instructor drives the performance vehicle on the track while providing instruction as the novice observes and listens. This instruction is typically auditory; however, may also include visual explanation via pointing. Thus, our explanatory communications are framed within an observational learning context, and explanations are motivated by these explanatory modalities. In safety-critical domains like driving, observational learning may be a safe and effective method to transfer knowledge to a novice. This study was designed to assess whether observing an AI coach drive and explaining its behavior are effective methods to improve a novice's driving performance, among other outcomes.

*Dimension 1: Information type*

First, we focus on the type of information presented to a driver: 'what' information provides descriptors of where the vehicle should drive, specifically following the ideal racing line; 'why' information explains why that position and movement is optimal. These are inspired by the explanations examined in experimental work on explanations of driving behavior by Koo et al.[26]; discussed in frameworks by Wang et al.[15] and Lim et al.[43]; and explored in cognitive science and philosophy[44]. In our experiment, the 'what' explanations are designed to help the participant know where to go on the course in order to follow the racing line, such as "to the left edge of the track." The 'why' explanations are designed to help the participant develop a more generalized understanding of performance-driving concepts, techniques, and decision-making. These include ways to optimize speed, acceleration, and lap time. For example, the vehicle moves "to the left edge of the track ...*to decrease the traction needed to get around the curve.*" We separate participants into 'what'-only and 'what + why' conditions and compare these to a *no explanation* control group to assess the effect of information type.

*Dimension 2: Information modality*

We are further interested in information presentation modality. To assess the impact of multimodal explanations, two different 'what + why' conditions were tested. One received visual 'what' explanations, while the other received only auditory cues. This visual 'what' explanation shows a visual projection of the racing line (Fig. 2). Auditory 'what' explanations were designed to convey the same information given by the visual racing line instead of using verbal cues. Auditory 'why' explanation content remained consistent as the instructions conveyed were deemed too complex for simple visualizations.

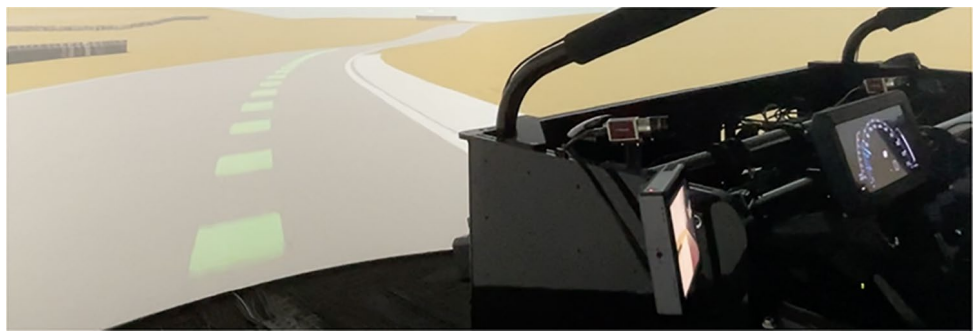

**Figure 2.** The green racing line projected on the track is an example of a visual 'what' explanation seen by Group 4.

### Conditions
Participants were randomly assigned to one of four experimental conditions (Table 1).

### Measures
*Driving performance*

1. *Racing line distance*: This is how far participants drove from the ideal racing line, measured in meters. It is calculated as the absolute value of the difference between a participant's line position and the ideal line position determined by the RL model, averaged across the track. Lower racing line distance means a participant was closer to the ideal.
2. *Maximum speed*: This is the highest speed achieved in miles per hour. Higher speed is indicative of better performance.
3. *Average Acceleration*: This is calculated as the mean change in velocity per 1 second when accelerating forward, measured as meters per second per second. Higher acceleration is indicative of better performance.
4. *Lap time*: The time taken to complete one lap in seconds. Lower time is indicative of better performance.

*Trust*

1. *Trust in AV driving coach*: This is the average of four, 5-point Likert scale (strongly disagree–strongly agree) questions similar to "I would trust the advice given to me by an AI driving coach."
2. *General trust in AVs*: This is the average of four, 5-point Likert scale (strongly disagree–strongly agree) questions similar to "I would trust an autonomous vehicle to drive me around safely."

*Expertise*

1. *Self-reported performance driving expertise*: This is the average of three, on a 5-point Likert scale (strongly disagree–strongly agree) questions similar to "I understand the concepts behind performance driving."
2. *Racing line knowledge*: This is the number of correct responses given on four true/false questions and one diagram question designed by the research team to test how well participants understood the racing line concept. It was given after all driving tasks concluded.

*Confidence and Cognitive load*

1. *Performance driving confidence*: This is the rating, from 0-10, given in response to the question, "If you were asked to performance drive in real life (without assistance), how confident would you feel?"
2. *Cognitive load*: This is measured using the widely accepted NASA Task Load Index (TLX)[45].

### Procedure
The study was approved for human subjects research by WCG IRB (external) and informed consent was collected from all study participants before procedures began. All study procedures were performed in accordance with relevant guidelines and regulations. Figure 3 shows the study procedure.

The study was framed in the context of learning performance driving. Before and after exposure, participants drove two laps. First, a practice to familiarize participants with the track, controls, and vehicle capabilities. Next, a performance where participants were told to drive as fast as possible without going off track or losing control of the vehicle as if they were going for their best lap time. These instructions align with the standard instructions for racing used in past studies[5].

During the observation, participants observed the AI coach drive four laps around the same track in the driving simulator. During the first and last lap, the RL agent drove at full speed; these did not contain any explanations from the AI coach and remained consistent across all groups. The second and third laps occurred at a slower speed and contained explanations from the AI coach. The same explanations were repeated twice to promote uptake: in all, about 8 min of explicit instruction was given to participants. Explanations varied by the condition the participant was assigned. Participants in Group 1 did not receive any verbal or visual instruction during these laps but still observed the vehicle drive the course.

| Group 1: No explanation-Control (n = 10) | |
|---|---|
| Group 2: Auditory 'What' (n = 10) | |
| Group 3: Auditory 'What' + Auditory 'Why' (n = 11) | Group 4: Visual 'What' + Auditory 'Why' (n = 10) |

**Table 1.** Experimental conditions. Visual 'What' Example: A visualization of the racing line on the track. Auditory 'What' Example: "Move the car to the left edge of the track." Auditory 'Why' Example: "Moving to the left will widen the arc of the next turn, preserving momentum".

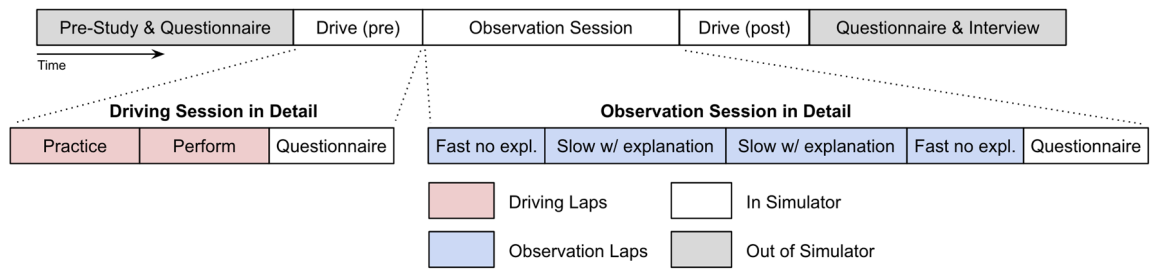

**Figure 3.** Study timeline. The study duration was approximately 1 hour in total and included a questionnaire, participant driving, AI coach observation, and interview.

Questionnaires in the simulator were given for measures of confidence and cognitive load. Questionnaires outside of the simulator to track trust, and expertise, as well as to get feedback on participants' experience overall. A 10–15 min semi-structured interview was conducted at the end, during which participants gave feedback on the AI coach's instructions, described their own learning procedure, and discussed trust and opinions on autonomous vehicles and coaches (see Supplementary Information for more details).

## Results
### Analysis methods

To assess score differences from before to after observation, we used linear mixed effects (LME) models via the R package 'lme4'. LME models yield similar results as mixed model ANOVAs, however, they allow for greater flexibility for pre-post experiments while allowing random effects to reduce the probability of a Type 1 error[46]. To test the impact of the explanations received by different groups, models were made with fixed effects for Timepoint (pre-post) and Group (1, 2, 3, or 4), with random effects for Subject ID added to control for individual differences. This tests if the pre-post *change* experienced by each experimental group differs from the pre-post *change* of Group 1, the pure-observation control. In some cases, we are also interested in the impact of the observation sessions themselves, regardless of group assignment. This can give insight into the general impact of observing any version of the AI coach. In these cases, the fixed effect for Group is removed, allowing us to test pre-post differences with all groups combined. To avoid skewing results, outliers were removed from racing line analysis when they represented extreme values related to a participant's vehicle moving far off the track; other racing measures are still valid in these cases.

### Impact of AI coaching information type and modality on driving performance

A summary of driving performance results is included in Table 2 and Fig. 4.

Distance to the ideal racing line is our primary measure of AI coaching success, as vehicle positioning was the primary focus of the training. This is a measure unlikely to improve from practice alone due to its lack of intuitiveness for a novice driver. Looking between groups, we find significant differences in racing line distance. Specifically, differences are found between groups receiving explicit instruction (2, 3, 4) compared to the control group (1) who did not receive explicit instruction. All groups with explicit instruction saw favorable pre-post change (i.e. were closer to the ideal racing line), while the control group itself got worse pre-post. For Group 2 ($\beta = -0.3$, t(35) = $-2.3$, $p < 0.05$) and Group 4 ($\beta = -0.3$, t(35) = $-2.0$, $p < 0.05$), differences were significant compared to Group 1 (Fig. 4). These results imply that explicit instruction on the racing line was helpful for learning beyond simple observation and that both information content type and information modality play a role in how effective an AI coach will be.

| Measure | Δ Group (vs. Δ Group 1 Control) | Δ Pre-Post (groups combined) |
|---|---|---|
| Lap time | . (trending) Group 2 (+) | *** (−) |
| Max speed | ** Group 2 (−) | *** (+) |
| Racing line distance | * Group 2 (−), * Group 4 (−) | NS |
| Average acceleration | ** Group 2 (−) | *** (+) |
| Trust in AV coach | NS | *** (+) |
| General trust in AVs | NS | ** (+) |
| Self-report Perf. driving expertise | NS | *** (+) |
| Self-report driving confidence | NS | *** (+) |

**Table 2.** Results summary. Impact of AI Coaching measured via LME. Direction of effect is indicated in parentheses. Sig. Codes: '***' $p < 0.001$ | '**' $p < 0.01$ | '*' $p < 0.05$ | '. (trending)' $p < 0.1$.

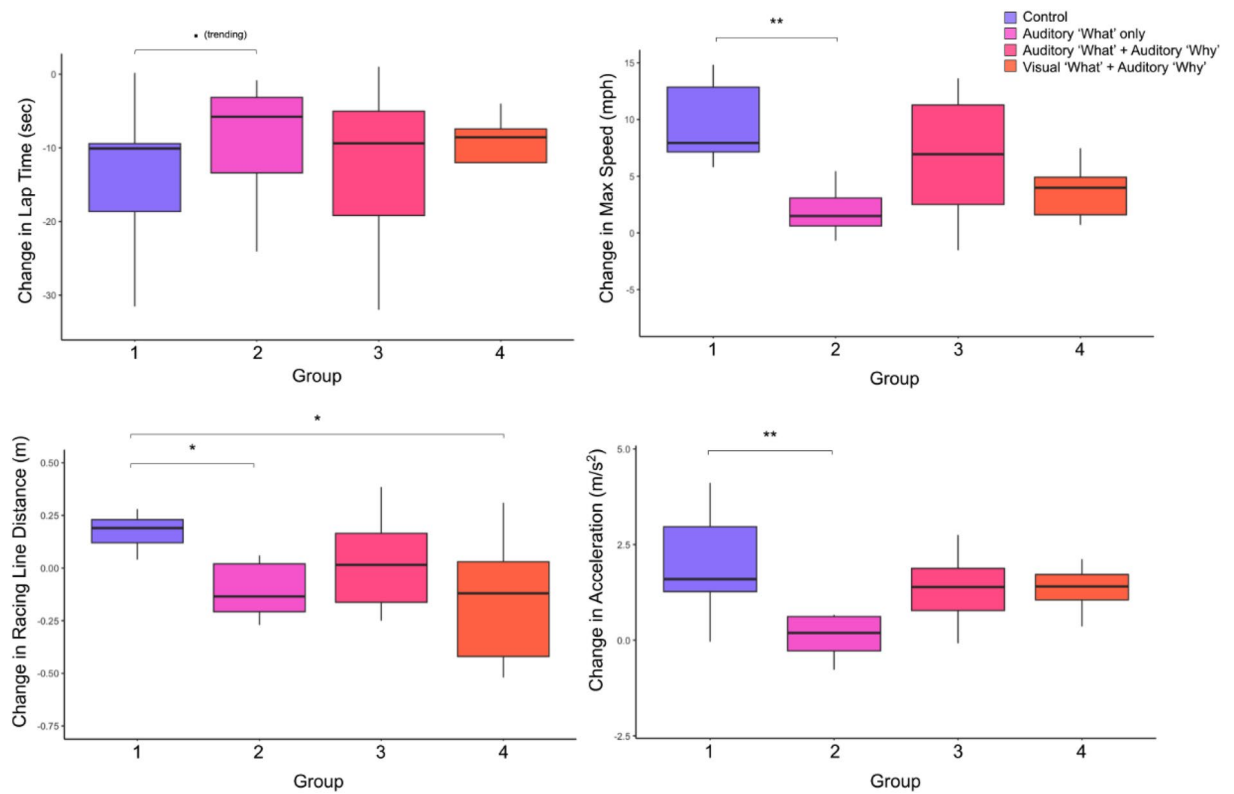

**Figure 4.** Change (Δ) from pre-post observation for driving performance measures by group.

Among other driving performance measures, Group 2 saw the most differences compared to Group 1. Group 2 improved less on lap time ($\beta = 10.3$, t(37) = 1.8, $p = 0.08$), max speed ($\beta = -9.0$, t(37) = $-3.4$, $p < 0.01$), and acceleration ($\beta = -1.6$, t(37) = $-2.7$, $p < .01$) compared to the pre-post improvement of the control (Fig. 4). Group 3 and 4 did not see any significant pre-post differences compared to the control by these measures.

Combining across all groups allows us to assess pre-post changes that stem from simply observing any version of the AI coach. When participants are lumped together, results show improvement from pre to post-exposure for several aspects of driving performance, including faster lap times ($\beta = -13.8$, t(40) = $-6.8$, $p < 0.001$), greater average acceleration ($\beta = 1.3$, t(40) = 5.9, $p < 0.001$), and greater max speed ($\beta = 5.7$, t(40) = 5.5, $p < 0.001$). Interestingly, we did not find an all-groups-combined benefit of the AI coach on a participant's average distance to the ideal racing line, further emphasizing the importance of the specific explanation received. These results as a whole demonstrate a clear positive trend on the impact of observing the AI driving coach in helping improve these aspects of driving performance. They also introduce nuance in the impact of different types and modalities of explanations on performance results.

### Impact of AI coaching information type and modality on AV trust, self-perceived confidence, and expertise

As a whole, impressions of the AI coach were very positive across all groups: 90% of participants agreed the AI coach was helpful, 93% agreed it helped them understand performance driving better, and 85% agreed it helped improve their driving.

Observing the AI driving coach was subjectively impactful for all groups with regard to trust, confidence, and self-reported expertise. We found all-groups-combined pre-post increases in perceived trust in driving coach ($\beta = 1.5$, t(40) = 5.6, $p < 0.001$) and in autonomous vehicles in general ($\beta = 0.9$, t(40) = 3.3, $p < 0.01$). Participants had higher self-report performance driving expertise ($\beta = 1.0$, t(40) = 3.7, $p < 0.001$) and confidence in their performance driving skills ($\beta = 1.4$, t(40) = 6.6, $p < 0.001$). We did not find significant differences in these measures between groups, however. See Table 2 and Fig. 5 for a summary.

Scores on the 5-question racing line knowledge quiz—which was only given post-observation—showed that Group 4 trended towards being higher than Groups 1 ($p = 0.06$) and 2 ($p = 0.10$), but did not reach statistical significance. These differences were assessed using ANOVA with a Tukey HSD post-hoc test.

We assessed if any of these measures-trust, confidence, or expertise—were able to predict a participant's pre-post change in racing line distance. This was done by categorizing participants into groups based on score percentile. We found participants in the top 33% percent of trust in coach ($\beta = 0.4$, t(36) = 3.1, $p < 0.01$) and trust in AVs in general ($\beta = 0.3$, t(35) = 2.2, $p < 0.05$) moved closer to the racing line compared to those in the bottom 33% of each trust measure respectively. Confidence, self-report expertise, and expertise measured via knowledge quiz did not appear to impact change in racing line distance.

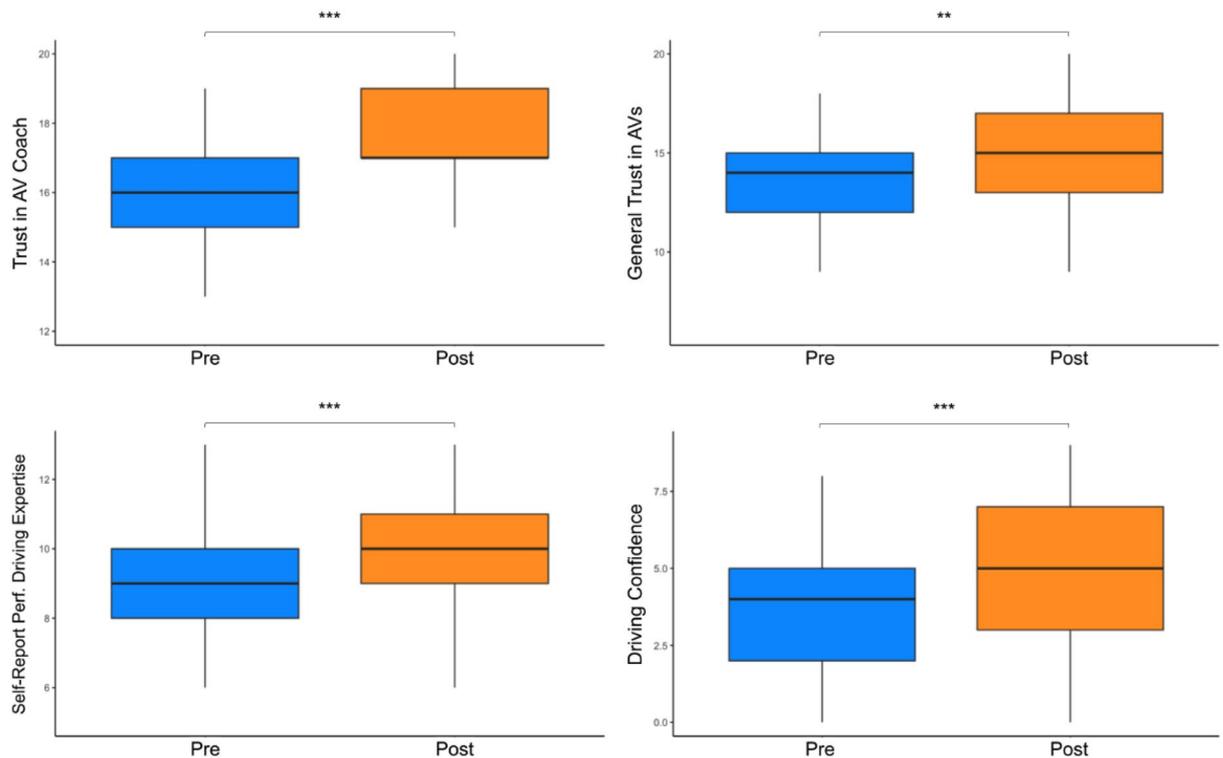

**Figure 5.** Pre-post scores for self-report measures, all groups combined.

### Impact of AI coaching information type and modality on cognitive load
We are interested in understanding the impact of each type of AI coach on cognitive load. Looking specifically at cognitive load directly after the observation sessions, we find that Group 3 had the highest cognitive load (mean = 14.2, se = 5.7), followed by Group 2 (mean = 14.0, se = 3.2), Group 1 (mean = 13.1, se = 7.0), and Group 4 (mean = 11.9, se = 5.4). Though not statistically significant from each other via ANOVA, these group scores follow the same trend reported in the interview phase and thus may be meaningful for explaining differences observed between groups.

We investigated if cognitive load before or after observation impacted pre-post change in racing line distance. We found individuals in the lowest 33% of cognitive load pre-observation improved less with respect to the racing line compared to those in the highest 33% ($\beta = 0.29$, t(37) = 2.5, $p < 0.05$). We also found those in the middle 33% of cognitive load post-observation got closer to the racing line compared to those in the highest 33% for the cognitive load ($\beta = -0.28$, t(36) = $-2.5, p < 0.05$). These imply that cognitive load impacts a participant's ability to learn and implement racing line knowledge.

### Interview insights
Qualitative analysis of the post-observation interview revealed insights into the learning processes of participants, how the HMI helped or hindered these processes, and brought a new understanding of participant dispositions towards trusting autonomous vehicles. Key themes emerged from the interviews that may inform future HMI design for AV Coaches.

*Information content, presentation, and modality*
**All groups wanted explanations of additional types.** For participants in Group 1—who received no explanations-explicit 'what' and 'why'-type information, such as those given to other groups, were desired by nearly all participants. Though many noticed patterns, participants experienced a large amount of uncertainty around what the AI Coach was trying to teach them. One participant explicitly stated, "at first … I didn't get what the [AI coach] was teaching me." Participants in Group 2—who received auditory 'what' information only— wanted a further rationale for the 'what' instructions they received (i.e. they wanted 'why' information). One participant commented that they "didn't know why the vehicle [moved in a particular way], so [they] couldn't generalize." Participants in Groups 3 and 4 expressed a desire for additional visual explanations to help them with the timing and magnitude of accelerating and braking inputs. Finally, a few members of Group 4 suggested that real-time or post-drive feedback would also be helpful for their improvement.

**Too much auditory information caused a feeling of being overwhelmed and a desire for more efficiency.** Participants in Group 3—who received auditory 'what' and auditory 'why' information—chiefly expressed a desire for *more efficiency* in the information presented to them. This was the result of Group 3's tendency to be overwhelmed with the amount of information presented. "Substantial", "overwhelming", and "a lot" were all used to describe the amount of information presented. In essence, participants still wanted comprehensive instruction covering a wide range of racing topics, but they wanted this instruction to be easier to consume and delivered in a less burdensome way. The rationale for this request was that it was too difficult to attend to and comprehend all of the auditory 'why' information while also receiving auditory 'what' information (though the two never directly overlapped).

**No explanation and auditory 'what' explanations introduced uncertainty.** In addition to uncertainty about what the coach was trying to teach them, Group 1 participants were unsure of exactly how to position themselves on the track or why they should do so. "It was hard to know what the important parts are [with observation alone]," noted another participant. This introduced additional challenges to the learning process. Participants in Groups 2 and 3 expressed difficulty with the preciseness of auditory 'what' instructions, as they had to internally map auditory visuospatial cues such as "move to the left edge" to precise visual positions on the track with inherent uncertainty. These participants wanted higher specificity in the position information they received in order to reduce the amount of guesswork.

**Visual 'what' information was preferred or desired near-universally.** Many participants in Groups 2 and 3 explicitly requested a visual racing line. For example, one participant expressed a desire for a "visual line showing … details such as how much to go left, etc." This was primarily as a means to overcome the uncertainty of auditory 'what' explanations. Group 4—who received visual 'what' and auditory 'why' information—appeared far more satisfied with their explanations than the rest. They expressed that the visual projection of the racing line was helpful, and in contrast to Group 3, they did not report any issues with overwhelm nor a desire for more efficiency. One participant commented, "the path on the track was … very helpful. [It] helped me feel more comfortable and confident".

*Trust in autonomous vehicles*
For nearly all 41 participants, a lack of trust in autonomous vehicles was a key issue hindering their willingness to adopt AV technology. Aligning with prior research on AV trust, the most common trust concern for our participants was over the AV's ability to perform safely and reliably[47]. There was a variance between participants: while some only expressed concerns for extreme or rare circumstances, others felt less confident in AVs for any circumstance involving the potential for unexpected occurrences (such as animals), pedestrians, or other drivers. This implies that trust in AVs may be contextually dependent and subject to individual differences. Other concerns included a lack of trust in the companies building the AVs, concerns over AVs running under-tested beta software, legal liability, losing the fun of driving, and AVs taking jobs from humans.

To help alleviate many of these concerns, participants reported that trust could be built through repeated, positive experiences with AVs. Many also suggested that AI explainability would help them feel more comfortable. Specifically, validation that an AI "perceives" what is around it, knowing how an AV will behave in specific driving situations, and understanding the rationale for decisions. Participants wanted to know *when* an AV can be relied upon and to maintain the ability to regain control from an AV if they are feeling uncertain or uncomfortable.

## Discussion

This study aimed to assess the viability of an AI driving coach for performance driving instruction as well as provide insight into how HMIs for AI driving coaching can be improved in the future. Using a mixed-methods approach (n = 41), we find that an AI driving coach may be a successful method to improve driving performance. Important considerations must be made to the type and modality of information presented. Our results shed light on how to design effective human-machine interfaces (HMIs) for coaching and driving interactions more generally. We discuss key findings and their implications for future HMI design in these contexts.

**AI Coaching is a viable method for performance-driving instruction.** Both quantitative and qualitative results of this study support the viability of an AI coach for performance-driving instruction. The primary area of focus for our AI coach was to provide instruction on racing line positioning. For racing line distance, we observed that Group 1 (control) got worse pre-post, while Groups 2 (auditory 'what' only), 3 (auditory 'what' and auditory 'why') and 4 (visual 'what' and auditory 'why') improved. The benefits of explicit instruction on racing line distance are unsurprising, as positioning on the racing line may be counterintuitive for novices. For instance, the racing line often follows the edge of the track, whereas everyday driving generally requires sticking to the middle of a lane. We also found significant overall pre-post-observation differences in several areas of performance driving, including faster lap times, max speeds, and acceleration rates before and after observation. Survey and interview data support AI coaching as an instructional method: most participants found the coach helpful, and overall boosts were found in participant expertise, trust, and confidence.

Proving nuance to our findings, Groups 2 and 4 were the most effective at improving racing line distance; however, Group 2 saw reductions in other areas—such as speed and acceleration—while Group 4 did not. This suggests that the specific instructions chosen greatly impact an AI coach's effectiveness. During the interview, each group expressed concerns and desires related to the way the type and modality of information impacted their learning. While these differed across groups, concerns and desires were generally shared by members of the same group. The fact that we see significant differences for Groups 2 and 4 but not for Group 3 further underscores the importance of carefully selecting the information conveyed by an AI coach.

We attribute group differences to how information directed attention, mitigated uncertainty, and influenced overload experienced by participants. These, in turn, affected how successfully participants could go about the learning processes.

**The type of information received by participants played a role in directing attention.** This aligns with prior work in explainable AI suggesting that a feature's presence (or absence) is meaningful and thus relevant for directing focus[16,44]. For our study, 'what' information aimed to teach participants how to adhere to the racing line. Group 1—who did not receive this information—suffered in this regard. Group 2 only received 'what' information, and thus, we suspect that they focused their attention more on adhering to the racing line and less on optimizing speed, acceleration, and lap time- topics discussed in the 'why' explanations. Thus, Group 2 saw less improvement in these other areas. By contrast, Groups 3 and 4 received 'what' and 'why' information, and thus focused their attention on several areas of performance driving simultaneously. Consequently, Group 4 also got closer to the racing line compared to the control group without seeing sacrifice in other measures like Group 2. We would have expected Group 3 to also have gotten closer to the racing line, as they also received the same types of information as Group 4. We suspect, however, that Group 3's improvement was hindered by two consequences of information modality: uncertainty and information overload.

**The uncertainty of auditory and visual 'what' explanations impacted the ease of processing.** For individuals in Groups 2 and 3, the uncertainty of auditory 'what' information presented to them was a major point of concern. To integrate 'what' information effectively, participants needed to transform auditory visuospatial cues such as "move to the left edge" into precise positions on the track. This extra step of transformation, which involves both creating a spatial representation for the linguistic cue and translating it into visual working memory[48], was not required with the visually presented racing line. Auditory 'what' cues were included in this study in alignment with Wickens' Multiple Resource Theory, which suggests that auditory information should be more efficiently incorporated during visually heavy tasks, such as driving, due to their different channels of processing[49]. In our case, however, we believe that presenting information auditorily made 'what' information more difficult to integrate due to the extra step of processing. For example, the instruction "move to the left edge of the track" lacks details specifying exactly *where* the participant should be aiming, and requires them to visually map the auditory spatial cue to a visual position on the track in front of them. This is supported by the modality appropriateness hypothesis of multisensory integration, which suggests that the effectiveness of information integration is dependent on the context of the task[50]. For a visual context of finding the racing line, visual cues were more efficient. Though we had expected the lack of explanatory preciseness for auditory explanations to be supplemented by observing the movements of the vehicle itself, participants expressed that the uncertainty of the instruction created too much "guesswork". As a result, our results support that visuospatial 'what'-type information should be presented visually—such as via a racing line projection—as opposed to auditorily.

**The modality and type of information affected the cognitive burden and overwhelm experienced by participants.** Decades' worth of prior research shows that increases in epistemic uncertainty or ambiguity can increase the cognitive burden of information processing[51,52]. Integrating these theories into our observations suggests that Groups 2 and 3 both experienced higher cognitive burdens processing the more uncertain auditory 'what' information than Group 4 did processing more precise visual 'what' information. For Group 2, who had no other information to pay attention to, this increase in demand may have been frustrating but did not impact their ability to stay on the racing line. For Group 3, we believe the increase in cognitive demand required to deal with uncertainty, combined with the increase in demand from receiving additional auditory 'why' information, was sufficient enough to overload Group 3 and prevent them from improving. In other words, there is a combinatorial effect of the burden of dealing with uncertainty and the burden of adding additional information to learn. Indeed, though not significantly different from the other groups, participants in Group 3 had the highest cognitive load and complained the most about being overwhelmed in their interviews. Group 4, who received visual 'what' and auditory 'why' information, had the lowest cognitive load of all groups. This aligns with prior work suggesting multimodal interfaces may promote more efficient processing of multiple sources of information[53], provided that each is integrated in modality-appropriate ways. Our theory is further supported by the Yerkes-Dodson law where too much cognitive burden is a detriment to performance[54]. We also found that Group 4's knowledge of the racing line trended towards being significantly higher than both Group 1 and Group 2, while Group 3 showed no difference between these non-why groups. This implies that the visual 'what' explanations and their lower cognitive burden may have helped Group 4's ability to take in auditory 'why' information compared to Group 3. Taken as a whole, these findings suggest that information modality and type influenced the amount of cognitive load and sensation of overwhelm for participants in each group, which affected performance and preference in turn. They also suggest that—in this particular instance—the uncertainty of information impacted the burden more than the sheer amount of information.

The implication of these results is that there is a nuanced relationship between information type and information modality for AI coaching HMI design. HMIs designed for AI coaching should aim to find a balance between comprehensive coverage of relevant topics and communication efficiency. This can help learners gain sufficient knowledge, prioritize attention, and avoid being overwhelmed. Context-based modality-appropriateness is essential for transferring information efficiently.

### Designing HMIs to support the learning process

The results of this study have clear implications for the future design of AI coaching and autonomous vehicle HMIs more generally. We briefly summarize eight design considerations.

**Designing for the learning process.** It is important to design for the specific learning process of the learners being taught. Combining participant descriptions with the insights provided by expert performance driving instructors allows us to delineate the process by which participants learned to drive in the driving simulator (Table 3). Prior work in explainable AI suggests that AI explanations that align with the learning or reasoning processes of their users may be more effective than those that are not[15,55]. Our study supports these past findings. By directing attention and supporting stages in the learning process differently—such as presenting information with more or less uncertainty or ambiguity—we found differences in the performance, knowledge, and preferences of our participants. Some participants expressed a desire for additional support with stages related to learning to brake and accelerate. The implication is that an AI coach, or HMI more generally, needs to be tuned specifically for the task at hand and that careful consideration should be placed on the process stages of learning that task. We delineated the learning processes of novices seeking to learn performance driving. Though not explored in this study, this presents the unique opportunity to study the effect of specific HMI designs on their effect of learning specific learning stages in future work.

| # | Stage | Description |
|---|---|---|
| 1 | Simulator calibration | In this phase, participants begin to calibrate the physical inputs on the simulator's steering wheel and pedals to the movement of the on-screen vehicle |
| 2 | Vehicle limits | Next, participants begin to learn what types of movements are possible from the vehicle. This often involves experimental speeding up, slowing down, and testing the grip of the tires, among other maneuvers |
| 3 | Track layout | Early in the learning process, this phase involves learning the different shapes the track may take. Later on, it may involve memorizing specific turns |
| 4 | Vehicle positioning | Combining phases 2 and 3 brings insight into how to position a vehicle to optimally move around turns. For example, moving to a specific edge. Finding this ideal racing line can be challenging and unintuitive for a novice driver, and was the specific target for the 'what' explanations presented |
| 5 | Handling in position | In this phase, participants build their handling skills by iteratively combining their understanding of vehicle positioning with their knowledge of the vehicle's limits. This is a mediary step between phase 4 with phase 6, and becomes increasingly important as the limits of speed and acceleration are tested. This was the first target for the 'why' explanations presented |
| 6 | Speed via Acceleration and Braking | In an ideal learning process, speed management should be learned in conjunction with positioning and handling. Once a participant understands where to be positioned, they can safely add and subtract speed to find an optimal balance. Adding speed without first understanding positioning can result in unsafe maneuvers, less ideal positioning, and- consequently- slower lap times. This was the second target for the 'why' explanations presented |
| 7 | Iteration | As drivers learn, they are able to build on their knowledge, iterate, and enhance specific motor skills, maneuvers, and racing strategies |

**Table 3.** Participant learning process phases. Though presented linearly, each sequential phase builds on the previous phases, which iteratively impacts the earlier phases in turn. Importantly, there may be individual differences in the learning process.

**Directing attention.** We find that-based on the type of information presented-attention may be directed in different ways. Omitting certain information, such as 'why' information in our study, may have caused participants to overly focus on the racing line to the detriment of other aspects of performance driving. The implication is that the specific information an AI coach presents should be mindfully chosen to ensure attention is being directed appropriately. In many cases, temporal ordering and prioritization of attention can be specifically designed for using techniques like scaffolding, organizing information hierarchically, and employing progressive disclosure so that less prominent details are deprioritized until they are needed.

**Balancing thoroughness with efficiency.** Carefully balancing information thoroughness with efficiency of presentation is crucial for useful HMIs. Increasing the amount of information conveyed to a novice may be helpful in transferring sufficient information, however, our evidence suggests that efforts will be futile if not done efficiently. As a result, careful selection of the type and amount of information presented is an important consideration for ease of processing. Avoiding overcomplexity and electing for easy to process information can help reduce the possibility of cognitive overload.

**Modality-appropriateness and minimizing uncertainty.** Our findings have clear implications on the importance of presenting modality-appropriate information that maximizes ease of processing. For driving instruction and HMIs for driving tasks, visual explanations are ideal for visual aspects of performance, such as showing where on the road to drive. Auditory may be more appropriate for information that is complex or not directly tied to a visual task, such as 'why'-type explanations. In both cases, information needs to be efficiently presented with emphasis on the precise details necessary for task execution. Presenting thorough, modality-appropriate information can minimize epistemic uncertainty and ambiguity.

**Trust as a barrier.** Trust was a major concern for our participants. Participants who trusted the AI coach more showed better performance for non-intuitive aspects of performance driving like following the racing line. As such, the effectiveness of an AI coach may be dependent on how trustworthy it is. The implications are clear and supported by a plethora of past research: without trust, HMIs will fail. According to our participants, trust can be built through repeated positive exposure, helpfulness, and explainability validating the AV's ability and reliability. Though coaching is a novel application from explainable AIs studied in prior work, similar principles for building trustworthy AI systems may apply. For the case of learning specifically, explanations that align with the learning and reasoning processes of the learners themselves may help build trust, as these give the learner the agency to cross-examine the information they are receiving in-context[15,55].

**Personalizing interactions.** Participants differed in their performance, preferences, and trust levels. Just as it is in human-to-human instruction, AI coaching will work best if personalized to the individual needs of the learner[56]. Attributes for personalization suggested from the study presented here may include ability, expertise, confidence, trust, preference, and cognitive load. For ability and expertise, an AI coach can add or withhold details and alter the complexity of information offered at a given time, among other techniques. For preference, individuals may want the type or modality of information offered to be changed based on their preferred learning style[57]. For trust and confidence, the types of supporting information given to a participant can be varied to address their specific trust concerns. These may include explanations of how the AI system will behave, how it was trained, or even reassuring comments on its ability. Designing for cognitive load may prove a larger challenge, but if an AV can detect when a participant is feeling overwhelmed through biometrics or self-report[58], it can modulate its communication style and behavior in commensurate ways.

**Contextual flexibility.** Particularly in discussions around trust, it was clear that certain contexts may require more or different information to mitigate concern than others. In this way, it is clear that pragmatics matter. This aligns with prior work on the impact of contextual factors on explainable AI usefulness[59]. For vehicle HMI design, this means that explanations may need to be altered based on the perceived riskiness or complexity of the driving scenario. In other cases, a vehicle may even give up control to a passenger, or vice-versa. For AI coaching, instructional explanations should be grounded in the context within which they should be applied, giving the learner a broader understanding of when certain lessons should be applied and when they should not be.

**Observation versus interactivity.** Our study highlighted the potential usefulness of observational AI coaching. We did not compare our observational methods to more interactive methods, such as question-answering[60]. Some participants requested real-time feedback and a means to interact with the information they received. These may be potentially viable future directions to explore in the context of AI coaching.

### Limitations and future work

This study was not without limitations. While these are limitations for a study that is designed to be a first step towards AI-driven vehicle coaching, this study still serves as a confirmation of the viability of AI coaching for driving and as a means to drive future work.

While between-group comparisons are the central focus of this research, all-groups-combined results allow us to assess the potential role AI driving coaches play in driving instruction more generally. Both group and combined comparisons, as well as qualitative results from interviews and surveys, support the viability of AI coaching as an instructional method. Combined results should be approached with caution, however, as it is not possible with the current study design to separate the pre-post observation changes from practice effects. Though changes from pre-post observation are far greater in magnitude than we would expect from practice alone, formally testing the impact of practice against the impact of training is left for future study.

We note that during real-world instruction, novices may ask questions to the instructor. Though the present study does not take into account this interaction, the explanations included were designed to cover the most common questions received by our expert drivers. Interactivity and other non-observational methods would be excellent future research directions to explore.

Compared to real-world instruction, our AI coaching sessions were quite short. We would expect larger differences and greater insight after exposure to longer AI coaching sessions and/or sessions that form a series.

Finally, though sufficient enough to obtain statistical significance and identify several data trends, this study had a relatively small sample size. With a larger sample, we would be able to evaluate interactions between variables such as cognitive load and trust directly on the pre-post differences in driving performance observed by the groups. We would expect to see clearer impacts of AI coaching with a larger study sample.

### Conclusion

In this study, we tested a novel use of AV technology-if AVs can teach humans to be better drivers. Results from the pre-post observation study reported here support the conclusion that observing an AI driving coach is a promising method to teach novices performance driving. Breaking participants into groups allowed us to determine how information type ('what' and 'why') and information modality (auditory and visual) influenced outcomes. We saw differences in how information directed attention, mitigated uncertainty, and influenced overload experienced by participants. These differences affected how successfully participants were able to learn. Results suggest that explanations for AI driving coaching and vehicle HMI design should seek to strike a balance between comprehensive coverage of relevant topics—such as how to follow the racing line and meet the limits of speed—with information complexity. When designed properly, explanations can direct attention to appropriate details efficiently, supporting the learning process while avoiding learner overwhelm. Context-based, modality-appropriate explanations should be opted for, especially when they mitigate information uncertainty. We conclude that communications must be designed to align with the needs and learning processes of the learner and present 8 design considerations to inform future HMI design that should generalize to driving in many different contexts, including everyday driving. These include specific suggestions for how to direct attention and choose the modality of an explanation as well as more general implications on the need for personalized, trustworthy, context-based HMIs.

### Data availability

The datasets generated and analyzed during the current study are available from the corresponding author upon reasonable request.

### Acknowledgements
The authors would like to express gratitude to the Human Interactive Driving and Human-Centered AI groups at Toyota Research Institute. A special thanks to Nadir Weibel, Allison Morgan, Andrew Best, Paul Tylkin, Tiffany Chen, Hiro Yasuda, Hanh Nguyen, and Steven Goldine for their individual contributions, feedback, and support.

### Author contributions
R.K., E.K., and J.C. conceived the experiment, R.K. conducted the experiment, and R.K. analyzed the results. All authors reviewed the manuscript.

### Competing interests
The authors declare no competing interests.

### Additional information
**Supplementary Information** The online version contains supplementary material available at https://doi.org/10.1038/s41598-024-62052-9.

**Correspondence** and requests for materials should be addressed to R.K.

**Reprints and permissions information** is available at www.nature.com/reprints.

**Publisher's note** Springer Nature remains neutral with regard to jurisdictional claims in published maps and institutional affiliations.

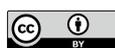